\documentclass[letterpaper]{emulateapj} 
\usepackage{apjfonts}                   
\usepackage{hyperref}
\usepackage{mathrsfs}
\usepackage{url}
\usepackage{lastpage}

\def\msun{{\,{\rm M}_\odot}}
\def\simlt{\lower.5ex\hbox{$\; \buildrel < \over \sim \;$}}
\def\simgt{\lower.5ex\hbox{$\; \buildrel > \over \sim \;$}}

\bibpunct[,]{(}{)}{;}{a}{}{,}

\begin{document}

\author{
Pau Amaro-Seoane\altaffilmark{1}\thanks{e-mail: Pau.Amaro-Seoane@aei.mpg.de},
Patrick Brem\altaffilmark{1}\thanks{e-mail: Patrick.Brem@aei.mpg.de}
\& Jorge Cuadra\altaffilmark{2,\,3}\thanks{e-mail: jcuadra@astro.puc.cl}
}

\altaffiltext{1}{Max Planck Institut f\"ur Gravitationsphysik
(Albert-Einstein-Institut), D-14476 Potsdam, Germany}
\altaffiltext{2}{Departamento de Astronom\'ia y Astrof\'isica, Facultad de F\'isica,
      Pontificia Universidad Cat\'olica de Chile,
            782-0436 Santiago, Chile}
\altaffiltext{3}{Max-Planck Institut f{\"u}r Astrophysik,
      Karl-Schwarzschild-Str.\ 1, D-85741 Garching bei M\"unchen, Germany}

\date{\today}

\label{firstpage}

\title{Tidal disruptions in circumbinary discs (I):\\
Star formation, dynamics, and binary evolution}

\begin{abstract}
 {In our current interpretation of the hierarchical structure of
 the universe it is well established that galaxies collide and merge
 with each other during their lifetime.
 If massive black holes (MBHs) reside in galactic centres, we expect them
 to form binaries in galactic nuclei surrounded by a circumbinary disc.}                                                           %
 {If cooling is efficient enough, the gas in the disc will clump and
 trigger stellar formation in situ.
 In this first paper we address the
 evolution of the binary under the influence of the newly formed stars, which form individually and also clustered.}
 We use SPH techniques to evolve the gas in the circumbinary disc and to
 study the phase of star formation. When the amount of gas in the disc
 is negligible, we further evolve the system with a high-accurate
 direct-summation $N-$body code to follow the evolution of the stars,
 the innermost binary and tidal disruption events (TDEs). For this, we modify the direct $N-$body code
 to (i) include treatment of TDEs and to (ii) include ``gas cloud particles'' that mimic the gas, so that the stellar clusters do not disolve
 when we follow their infall on to the MBHs.
 We find that the amount of stars disrupted by either infalling stellar clusters
 or individual stars is as large as $10^{-4}$/yr per binary,  higher than
 expected for typical galaxies.
\end{abstract}

\maketitle

\section{Introduction}
\label{sec.intro}

Super-massive black hole (MBHs) binaries are expected to form after major
galaxy mergers. The main driving mechanism for the MBHs to sink to the centre
is dynamical friction, where they will form a binary and start to shrink the
semi-major axis on their way to the final merger.  Slingshot of stars from the
surrounding stellar environment help the binary to further decay by exchanging
energy and angular momentum, down to distances of about 1 pc \citep{BBR80}. However, if the
amount of stars to interact with is depleted, there is a risk of stalling, so
that the MBHs would not coalesce within a Hubble time.  This is the so-called ``last-parsec problem''
 \citep[see][for a review on the whole process and references therein]{MM05}.

Key factors to surmount this last ``snag'' in the evolution
are, among others, the fact that (i) in the case of binaries with a total mass
of $\leq 10^7\,M_{\odot}$, slingshot ejections suffice to
guarantee coalescence within a Hubble time \citep{MM03}; (ii) the role of gas
may be crucial in the evolution of the binary, starting at larger scales. It
might well be that in a merger of gas-rich galaxies, if MBHs are present, they
will coalesce soon after the galaxies merge, in some $10^7$ Myr, if the gas is
distributed spherically. If, on the other hand, the gas is forming a
nuclear disc, the galaxies need only to have $1\%$ of their total mass in gas
for this to happen.  \citep{EscalaEtAl04,EscalaEtAl05}.
 \cite{CuadraEtAl09}
found that such gas discs could indeed commonly help in the merger of SMBHs
with masses in the range of our study, whilst this mechanism fails for masses
larger than $\sim 10^7\,M_{\odot}$; (iii) following with stellar dynamics,
resonant relaxation creates a steady state current of stars which can be as
large as ten times the non-coherent two-body relaxation
\citep{HopmanAlexander06}. This is a potential source of new stars populating
the depleted loss-cone; (iv) the work of \cite{BerczikEtAl06} shows that
considering a non-spherically symmetric system the final parsec problem is
largely solved (v) massive perturbers, such as giant molecular clouds or
intermediate-mass black holes, can accelerate relaxation by orders of magnitude
compared to two-body stellar relaxation, so that many new stars are supplied to
interact with \citep{PeretsAlexander08}; (vi) it has been observed that young,
compact star clusters such as the Arches and Quintuplet systems reside near the
Galactic centre. If these star clusters have masses larger than
$10^5\,M_{\odot}$, they can make their way down to the Galactic centre even if
they start from a distance as large as 60 pc within a few million years
\citep{McMillanPortegiesZwart03}. The tidal stripping of these young stars
could eventually provide the binary system with a new set of some $\approx
10^5$ stars; (vii) if intermediate-mass black holes (IMBHs), with masses ranging
between $10^{2-4}\,M_{\cdot}$ exist in the centre of clusters, it has been
predicted that within the innermost central 10 pc, we can expect to have some
50 IMBHs of masses $10^3\,M_{\odot}$, and still some of them at scales of a few
milliparsecs \citep{PortegiesZwartEtAl06}. The interaction of one of these
IMBHs with the binary of SMBHs would obviously accelerate the process of
shrinkage.

The studies just cited provide  a number of mechanisms to make the
binary shrink.  We expect then that a typical binary will be able to reach sub-pc separations,
especially in the case of relatively low-mass MBHs in gas-rich environments.
In this study, we concentrate on such a case (see section~\ref{sec:2mbhdisc} for details),
which is expected when the parent galaxies are gas-rich and large amounts of gas
 fall to the centre of the new system, together with the MBHs.  At
that location, the black holes get bound to each other, thus forming a binary,
and are surrounded by a massive, parsec-scale gaseous disc
\citep[e.g.,][]{EscalaEtAl05,MayerEtAl2007,DottiEtAl2007}.

Such gaseous discs are similar to proto-stellar discs:  due to their high
density compared to the central object tidal force, the discs will be locally
unstable to self-gravity, meaning that perturbations in their density field
will grow.  However, if the gas is unable to cool efficiently, then the gas
will not be able to contract and form clumps, and the density perturbations
will be sheared apart, creating a quasi-steady spiral pattern.  Remarkably, the
spiral pattern transports the angular momentum outward, making the disc behave
as an accretion disc.  On the contrary, if the gas is able to cool quickly
enough, then the density perturbations grow and form clumps, which shrink and
further accrete gas, breaking up the gaseous disc completely and turning it
into stars -- the so-called {\it fragmentation}
\citep[e.g.,][]{Gammie2001,RiceEtAl2005,NayakshinEtAl2007,Lodato2007,AlexanderEtAl2008,Paardekooper2012}.

In either cooling regime, the situation where at the centre of the
disc the central object is a binary will lead to a non-trivial
interaction between them.  On previous studies we have focused on
the inefficient-cooling regime, showing that torques
between the gas and the binary will shrink the orbit of the latter, while the
angular momentum is driven out through the disc
\citep{CuadraEtAl09,RoedigEtAl2011,RoedigEtAl2012}.
  In this paper we present the first
numerical study of the fast cooling regime in which the disc fragments into
stars, and follow the dynamical evolution of the binary--stars system.  We
carry out our study in two stages \citep[see also][]{KhanEtAl12}: first we model the fragmentation of the disc
using smoothed particle hydrodynamics, and then we switch to our
direct-summation $N-$body models to both follow the long-term evolution of the
system and to study the occurence of TDEs.

The reason for this two-step approach is that we first need to model
the gas hydrodynamics in order to follow the fragmentation process of
the gas, including the formation of stars and their growth via mergers
and accretion of gas.  In principle, one could wait for the gas to
disperse or be accreted, and simply continue the same integration to
follow the dynamical evolution of the stars for long time-scales.
However, the SPH code we are using is not designed to follow the
collisional N-body dynamics of the system, therefore, it is necessary
to use a different code that allow us to model the system of MBHs and
stars in a meaningful way.

\section{First stage in the evolution: Gaseous disc and star formation in situ}

\subsection{Two MBHs and a circumbinary disc}
\label{sec:2mbhdisc}

Following \cite{CuadraEtAl09}, we concentrate on a binary with the following
initial parameters:   total mass $M_{\rm bbh} = M_1 + M_2$, where $M_1$ and
$M_2$ are the masses of the individual MBHs, and mass ratio $M_1/M_2 = 3$, in a
circular (Newtonian) orbit of separation $a$.  The binary is surrounded by a
corotating gaseous disc with an initial mass $M_{\rm d} = 0.2 M_{\rm bbh}$ and
radial range $2a$--$5a$.  The gas is modelled as an ideal gas with $\gamma =
5/3$, and radiative cooling is mimicked with a cooling time defined as $t_{\rm
cool}(r) = \beta/\Omega(r)$, where $\beta$ is a free parameter that fixes the
cooling rate, $\Omega(r) = \sqrt{GM_{\rm bbh}/r^3}$ is the orbital frequency
around the binary, and $r$ is the distance from the binary centre of mass.
Since we are interested in the fragmentation regime, in this paper we consider
fast cooling rates, $\beta \le 5$. The choice of a disc that fragments is
 realistic for self-gravitating discs that cool thermally, above a
certain surface density threshold.  \cite{Levin2007} showed that, for the
masses and distances we are interested in here, that threshold lies in the
10--100 g/cm$^2$ range.

This model for the system dynamics is
scale-free, meaning that it can be scaled up or down to different masses and
lenghts.  However, in order to introduce star formation and also to estimate
the rate of tidal disruption events (TDEs), we need to choose physical units.
With that aim we set the total mass of the binary as $M_{\rm bbh} = 3.5\cdot \,
10^6\msun$ and we choose $a=0.04$\,pc.  This would be a typical mass for binary
black holes in the range that could be detected by a LISA-like experiment
\citep{Amaro-SeoaneEtAl2012,Amaro-SeoaneEtAl2012b}.
The chosen separation corresponds roughly to
the value where we would expect binaries to spend the longest of their
evolution in a simple model that considers binary shrinking due to stellar
scattering from a spherical cusp \citep{MM03} and torques
from a non-fragmenting disc \citep[see][ their eq.~12]{CuadraEtAl09}.
\footnote{Notice that the choice of $a = 0.04$\,pc is below the classical $\sim 1$ pc
separation of the ``final parsec problem'', but for the range of masses considered
in this work we deem it not a problem, as we summarized in the introduction.}

While several studies \citep{ASF06,Amaro-SeoaneEtAl09a,AS10a,PretoEtAl2011,KhanEtAl2011, KhanEtAl12} have shown
that stellar dynamical processes pump up the eccentricity of a binary
MBH, in this case we are assuming the binary has reached the inner
parsec in a gas-rich environment.  In such a case, the dynamical
friction of the gas on the MBHs drives them to form a circular binary
(e.g., \cite{DottiEtAl2007}). Thus we choose a circular orbit for the initial configuration.

\subsection{Implementation and treatment of the disc fragmentation}

To follow the process of circumbinary disc fragmentation, we use a modified
version of the smoothed particle hydrodynamics (SPH) code ({\sc Gadget},
\citealt{SpringelEtAl2001,Springel2005}), combining the numerical methods of
\cite{NayakshinEtAl2007} and \cite{CuadraEtAl09}.  Here we only briefly
describe the methods, and refer the interested reader to those papers for more
details.  We model the gaseous disc as an ensemble of initially $\approx
2\times10^6$ particles of $\approx 0.35\msun$ each. The code calculates the
gravitational and hydrodynamical interaction between gas particles, plus the
gravitational interaction between all particles, including the MBHs as well as
the ``proto-stars'' and ``stars'' that form during the simulation (see below).
We use a softening of $0.001a$ for the gas particles and of $0.01a$ for the proto-stars. The MBHs
do not use softening, but a sink radius within which gas particles are accreted. This radius had
a value of $0.3a$.

As initial conditions, we take the initially-circular system modelled by
\cite{CuadraEtAl09},
at a time $T \approx 500 \, \Omega_{0}^{-1}$.  In this
way we skip the transient initial evolution caused by the homogeneous initial
conditions described in their work, and start from a steady-state configuration
in which the circumbinary disc has developed spiral arms. Notice, however,
that their simulations used $\beta=10$, avoiding fragmentation.  In our
new simulations we set the value of $\beta$ to either 1, 2, 3, or 5.  As a
result, the disc now forms clumps, which grow in a runaway fashion.  Treating this
this with a pure SPH model is not feasible, as the growing densities require ever
shorter time-steps.  To circumvent this problem, we introduce sink particles to
model the proto-stars that we expect would form in these large density regions.

{\it Proto-star particles} are created when the gas density reaches 30 times
the Roche tidal limit, $M_{\rm bbh}/(2 \pi r^3)$.  How many stars will form out of a gas
density peak is a very complex question, whose solution is well outside the possibilities of
our study.  In our
model we deal with this issue  in an individual particle basis, i.e., each gas
particle is turned into one proto-star particle of the same mass.  However, the
newly formed proto-star particles can merge with each other, thus forming
higher mass stars. The merger criterion is simply that their distance is
smaller than $2 f_{\rm m} R_{\rm p}$, where $R_{\rm p}$ is the size of the
proto-stars, which we typically take as $10^{15}\,$cm, and $f_{\rm m}$ is a
free parameter with fiducial value of unity
that mimics the effect of gravitational focusing.  The size
parameter corresponds to $\sim H/10$ (where $H$ is the disc scale-height),
which is roughly the thickness of the gas arms we observe in the simulations.
Thus, in a two-step process we are in principle allowing all the dense gas
within the same overdensity to form one proto-star.  However, we only allow the
proto-stars to merge with each other as long as their masses do not exceed
$30\msun$.  Once they reach this mass we turn the proto-star particle into an
actual {\it star particle}.  The motivation for this threshold is twofold:
numerically, we form an actual star out of $\simgt100$ gas particles;
physically, we avoid the rapid formation of extremely massive stars.  Stars can
merge with proto-stars, but not among each other.

Stars and proto-stars also grow by accreting their surrounding gas.  We use an
Eddington-limited Bondi--Hoyle prescription to calculate their accretion rate,
and then pick up at random enough particles from the (proto-)star neighbours
that are merged with the sink particle \citep{SpringelEtAl2005}.  To calculate
the Bondi--Hoyle and Eddington accretion rates, we use the mass of the (proto)
star, and a radius that is either the main sequence value corresponding to that
mass \citep[eq.\ 11 in][]{NayakshinEtAl2007} for the star particles, or the
fixed value $R_{\rm p}$ for the proto-stars.  This difference results in a much
faster growth for proto-stars than for stars.

The black holes also accrete the few gas particles that get too close to them.
This procedure is done mostly to avoid the short time-steps that would be
required to follow those gas particle orbits.  Accretion on to the black holes
is modelled simply with a sink radius -- all gas particles entering the region
around $0.3a$ of either black hole are taken away from the simulation, with
their mass and momentum being added to the corresponding MBH \citep{CuadraEtAl2006}.

We have ran 6 different SPH simulations.  Four of them use the fiducial values
mentioned above, but differ on the strength of the cooling.  We refer to these
runs as {\tt beta1},  {\tt beta2},  {\tt beta3a} and  {\tt beta5}.  Additionally, since we
tend to form many very massive stars, we explore the effect of decreasing the
numerical size of the proto-stars, hindering their growth.
For $\beta=3$ then we run two additional simulations, {\tt beta3b} and
{\tt beta3c}, in both of which we use a smaller size for the protostars
of $1436.8\,R_{\odot}$ instead of the fiducial value of
$14,368\,R_{\odot}$.  Run {\tt beta3c} has however a larger gravitational
focusing factor of $f_{\rm m}=10$ instead of the fiducial $f_{\rm
  m}=1$.  For both extra simulations then there is a more severe
(Eddington) limit on the accretion rate for the proto-stars than in
the fiducial {\tt beta3a}, while simulation {\tt beta3b} has additionally
a smaller likelihood of proto-stellar mergers.

These choices in the conditions for gas cooling and for transforming
gas particles into ``stars'' arguably capture a sufficiently broad
number of potential fragmentation scenarios so as to envisage our analysis
representative of a self-gravitating disc, within the limitations of the
rather expensive numerical experiments.

\begin{center}
\begin{figure*}
\resizebox{0.94\hsize}{!}
         {\includegraphics[scale=1,clip]{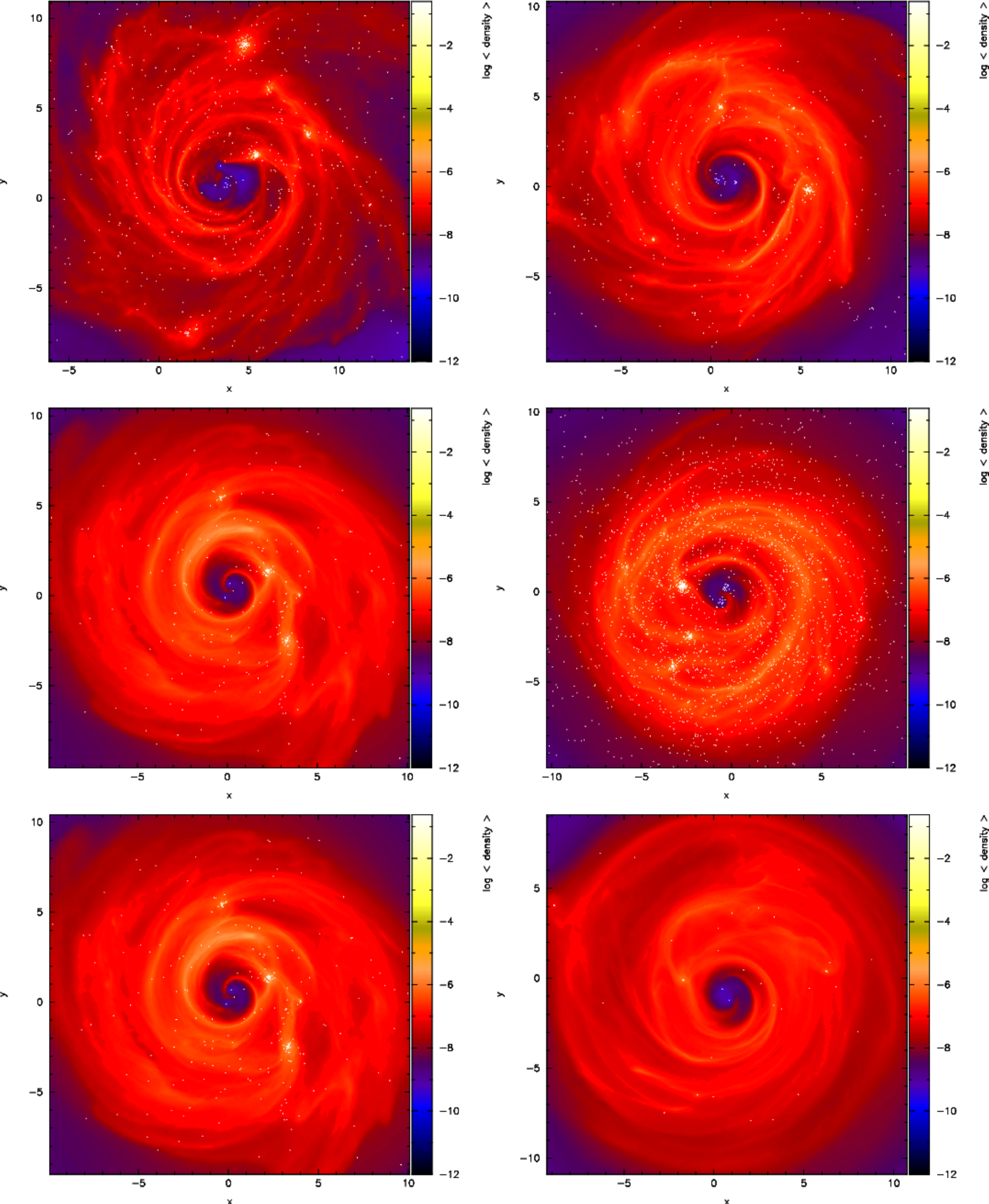}}
\caption
   {
Gas density projected in the X--Y plane perpendicular to the angular momentum vector of the
system for
{\tt beta1}, {\tt beta2}, {\tt beta3a}, {\tt beta3b}, {\tt beta3c} and {\tt beta5},
from the left to the right and from the
top to the bottom. White dots represent the ``sink'' particles, i.e. the MBHs
and the stars formed during the simulations. All snapshots are at $T =
300\,\Omega_{0}^{-1}$ but for the last one, which was integrated up to $T =
1000\,\Omega_{0}^{-1}$, because in that run cooling is quite slow and the number
of stars is still very low at earlier times (see Fig.~\ref{fig.Stellar_Mass}).
Note that there is virtually no difference betwen {\tt beta3a} and {\tt beta3c}.
The figures were made with {\sc Splash} \citep{Price2007}.
   }
\label{fig.GasDens_All}
\end{figure*}
\end{center}

\section{Fragmenting discs}
\label{sec.fragm}

We run the SPH simulations of circumbinary discs for several hundred
binary dynamical times.  Due to the gas self-gravity, clumps grow in
the disc.  Given the short cooling times, these clumps contract,
achieving the disc fragmentation.  In most simulations, after only
$\approx 200\,\Omega_{0}^{-1}$, the vast majority ($\simgt 90$\%) of
the gas is turned into stars, as expected.  The system then reaches a
quasi-steady state in which stars very slowly accrete the tenous
left-over gas \citep[see][]{NayakshinEtAl2007}.  The gas morphology at
that stage for the different simulations is shown in
Fig.~\ref{fig.GasDens_All}.

\begin{figure}
\resizebox{\hsize}{!}
          {\includegraphics[scale=1,clip]{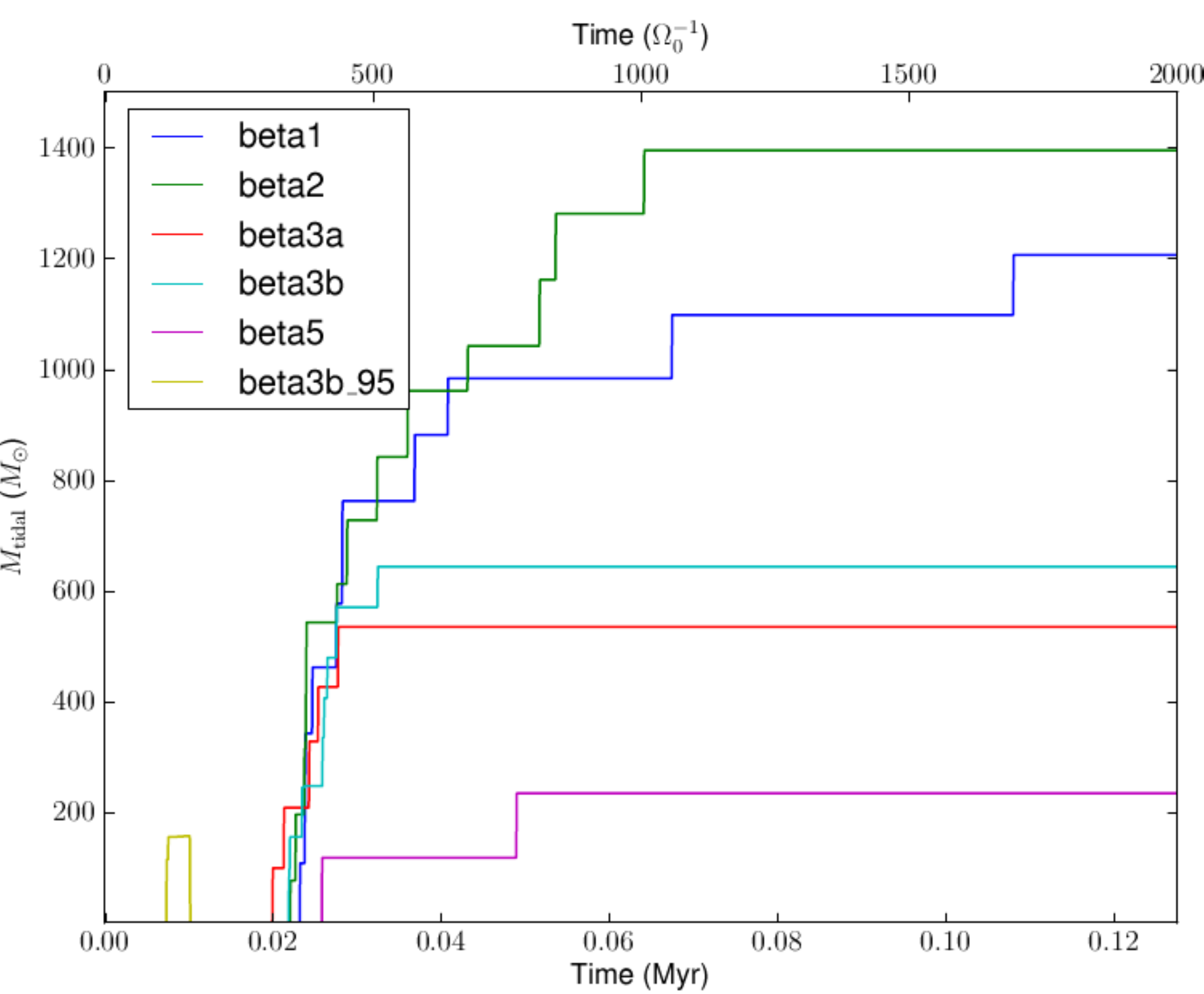}}
\caption
  {
Accumulated stellar mass formed in the disc in $M_{\odot}$ for our fiducial
case of a binary of $3.5\cdot 10^6\,M_{\odot}$. All simulations but for
$\beta=5$, which needs a bit longer, reach relatively fast the maximum of
stellar mass and saturate with values below $10^6\,M_{\odot}$.
   }
\label{fig.Stellar_Mass}
\end{figure}

The fragmentation rate is set by the cooling time of the disc, thus
discs with lower values of $\beta$ will evolve faster. We can see this
in figure \ref{fig.Stellar_Mass}, which shows the mass in stars as a function
of time for all the simulations. The fourth column in
Table~\ref{tab.nbody} shows the number of stars formed in each
simulation.  Considering only the variation of $\beta$, it is clear
that shorter cooling times result in larger amounts of stars, as
expected \citep{NayakshinEtAl2007}.  As the total stellar mass is
approximately constant, the typical stellar masses will be lower for
shorter cooling times.

It is interesting to note that the star formation process is not
uniform.  Instead, it happens preferentially in a few localised,
relatively large regions, whose sizes are set by the spiral-arm
overdensities.  Even though we allow proto-stars to merge when they
form close together, our numerical recipe avoids the formation of very
large stars, which forces the formation of ``stellar clusters'' (see
the left panel of Fig.~\ref{fig.3b_mosaic})
 \footnote{
      For a movie of this simulation, visit the URL\\
                \url{http://members.aei.mpg.de/amaro-seoane/fragmenting-discs}.\\ The encoding
                 of the movie is the free OGG Theora format and should stream automatically
                 with a gecko-based browser (such as mozilla or firefox) or with chromium or
                 opera.  Otherwise please see e.g.
                 \url{http://en.wikipedia.org/wiki/Wikipedia:Media_help_(Ogg)} for an
                  explanation on how to play it.}.
Some of these clusters feel a strong torque from the spiral arm
and are driven towards the centre of the system, where the tidal force
of the binary disperses them.  This stellar distribution affects the
long-term dynamics of the system and has
interesting consequences for the production of tidal disruption events
(\S \ref{sec.ClusterInfall}).

In our tests with $\beta=3$ and different stellar growth recipes we
first notice that runs {\tt beta3a} and {\tt beta3c} are practically
identical, and that run {\tt beta3b} has the same curve of stellar mass
growth.  From this we conclude that in our simulations accretion is
not important and that stellar growth is driven by mergers of sink
particles. \footnote{This is actually not surprising, as stars grow by
  mergers in roughly the dynamical time inside an overdensity, $t_{\rm
    dyn} \sim (G \rho)^{-1/2}$, which corresponds to about hundred
  years for the density values required for the introduction of sink
  particles.  On the other hand, the Bondi accretion rate for a solar
  mass sink, even for those very high densities, is only $\dot M_{\rm
    Bondi} \sim 10^{-5} M_\odot {\rm yr}^{-1}$ in our models, so the
  time required to accrete {\it a single} SPH particle turns out to be
  $t_{\rm acc} \sim 3\times10^4\,$yr.}  We also notice that the number of stars formed is
about an order of magnitude higher in {\tt beta3b}, which has 10 times
smaller proto-stars than the fiducial run, and that the effect of
having smaller proto-stars in the simulation is similar to having a
shorter cooling time.

To continue our study of the evolution of the MBHs and circumbinary
disc system, we will take the masses, positions, and velocities of all
sink particles and use them as input in direct-summation $N-$body
simulations.  For simplicity, we take the snapshot at time
$T=300\,\Omega_{0}^{-1}$ for all configurations, except for {\tt beta5}.
Since in that run the evolution is slower, we use the snapshot at
$T=1000\,\Omega_{0}^{-1}$, by which time 90\% of the gas has turned
into stars.

\section{The role of stars in the shrinking of the binary}
\label{sec.shrinking}

To analyse the dynamical evolution of the  MBH binary embedded in the stellar
system product of the stellar formation we use a direct-summation code, {\sc
Nbody6}.  This is a very expensive method because we integrate all
gravitational forces for all formed stars at every time step, without making
any a priori assumptions about the system. This code belongs to the family of
dynamical codes for particle systems with relaxation processes of Sverre
Aarseth. The code uses the improved Hermite integration scheme as described
in~\citep{Aarseth99,Aarseth03}.  Since these approaches integrate Newton's
equations directly, all Newtonian gravitational effects are included naturally.
More crucial for this subject is that it also incorporates both the
{\em KS regularisation} and the {\em chain regularisation}, so that when stars
are tightly bound or their separation becomes too small during a hyperbolic
encounter, the system is regularised \citep{KS65}. The advantages of this code
as compared to the leap frog integrator of {\sc Gadget} for our particular
problem are obvious, namely the high accuracy in the energy conservation, since
we are interested in the correct evolution of the inner binary of MBHs as well
as in potential TDEs. For this aim, as we describe later, we modified the
standard version of {\sc Nbody6}.

For each simulation, the initial masses, coordinates and velocities for the
stars and MBHs are taken from the {\sc Gadget} data at the times shown in
table~\ref{tab.nbody}.  At that moment, the gas mass -- stellar mass ratio is
very low (see table \ref{tab.nbody}, column $M_{\rm gas}/M_{\star}$). The
gravitational effect of gas is almost negligible and we do not include it in
the simulations.  Despite our limit to the growth of ``proto-star particles''
in the SPH simulations (see section~\ref{sec.fragm}), some ``star particles''
did manage to achieve very large masses.  We deem those unphysical, so in the
initial conditions for our $N-$body runs we replace  stars with masses $m$
above $120\,M_{\odot}$ with a cluster following a Plummer distribution
\citep{Plummer11} consisting of equal mass stars with total mass $m$ and radius

\begin{equation}
   R = \Bigl(\frac{m}{3 M_{\rm bbh}}\Bigr)^{1/3}\,r,
\end{equation}

\noindent
with $r$ the distance to the centre-of-mass of the binary.
The last equation corresponds to the Roche lobe of the massive star with respect to the MBH
binary with mass $M_{\rm bbh}$.

\begin{figure}
\resizebox{\hsize}{!}
          {\includegraphics[scale=1,clip]{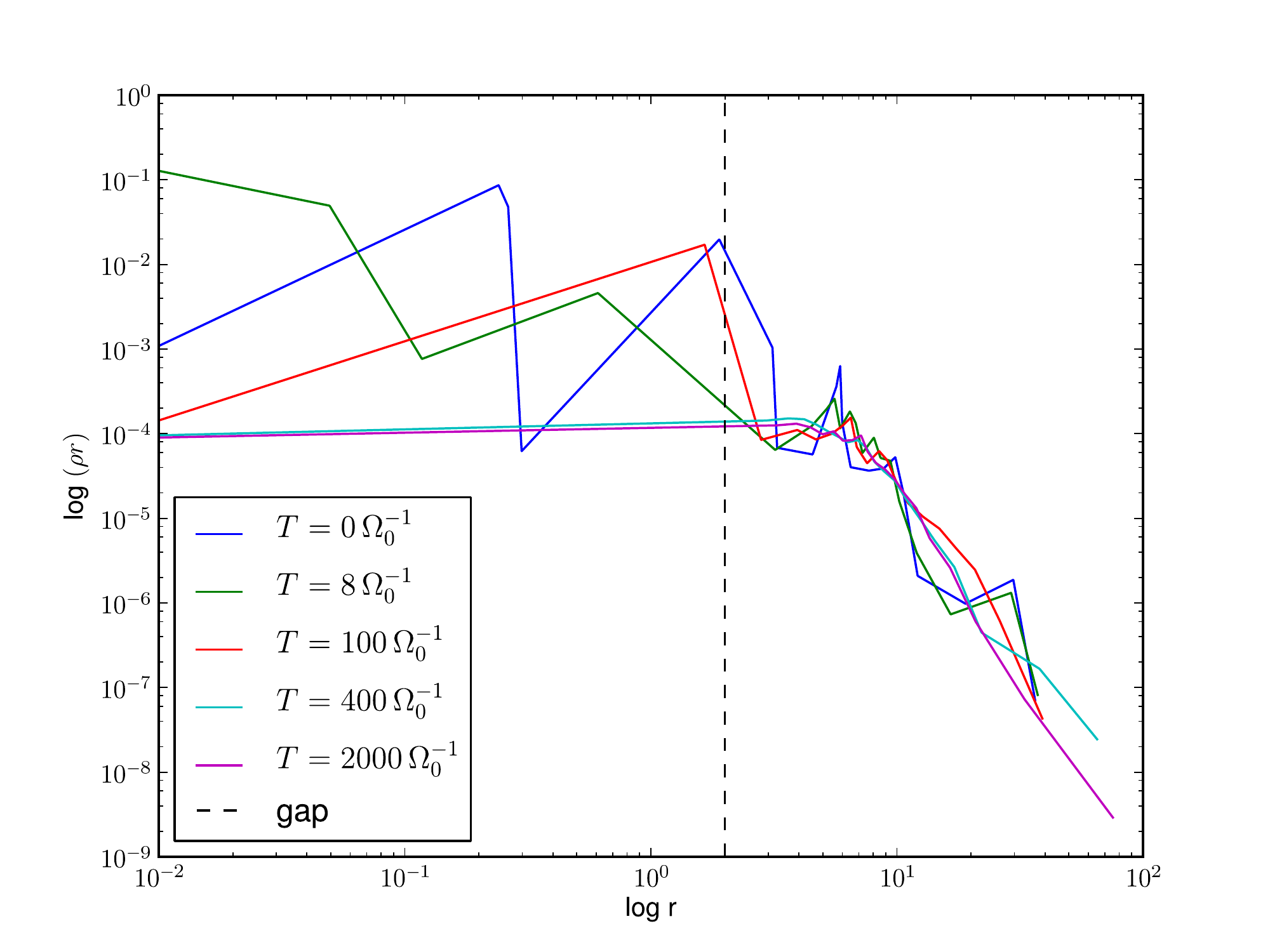}}
\caption
   {
Evolution of the density profile in one $N-$body simulation {\tt beta1}
at different times in the evolution. The dashed line corresponds approximately to the
position of the inner gap in the SPH simulation.
   }
\label{fig.profile_evolution}
\end{figure}

In our $N-$body simulations, table \ref{tab.nbody}, we exclude stars which are
at a distance $r > 100 a$, where $a$ is the semi-major axis of the MBH binary.
We assume those stars would have only a negligible effect on the binary
evolution.  They correspond to about a quarter of all stars in each simulation.
As shown in figure \ref{fig.profile_evolution}, this cut in the
cluster did not affect its global structure, its density profile remains roughly constant at large radii.
The figure also shows that the region
inside a few times the binary semi-major axis gets depleted quickly by sling-shot
interactions, as expected.

\begin{table}
  \centering
    \begin{tabular}{ | l | l | l | l | l | l |}
      Model & {SPH} time & $M_{\rm gas}/M_{\star}$& $N_{\rm SPH}$ & $N_{\rm NB}$ & $N_{\rm split}$ \\ \hline
      {\tt beta1}    & 300  & 3\%  & 2536 & 1895 & 4469  \\ \hline
      {\tt beta2}    & 300  & 7\%  & 1429 & 1141 & 2768  \\ \hline
      {\tt beta3a}   & 300  & 9\%  & 699  & 585  & 1924  \\ \hline
      {\tt beta3b}   & 300  & 9\%  & 5487 & 4486 & 5193  \\ \hline
      {\tt beta5}    & 1000 & 10\% & 167  & 144  & 1146  \\ \hline \hline
      {\tt beta3b95} & 95   & --   & 5540 & 5540 & 5540  \\ \hline
    \end{tabular}
    \caption{
    Initial data for the {\sc NBODY6} runs.
Notice that we do not integrate run {\tt beta3c} using the $N-$body technique, because it turned out to be identical to {\tt beta3a}.
    {SPH} time is the moment at which
    we stop the {\sc Gadget} simulation, in units of $\Omega_{0}^{-1}$, $M_{\rm
    gas}/M_{\star}$ is the ratio between gas and stellar mass at that moment, $N_{\rm
    SPH}$ is the number of stars that have been formed at that moment in the
    {\sc Gadget} simulation, $N_{\rm NB}$ is the
    number of stars within a distance $r < 100 a$ from the centre
    of mass of the binary and $N_{\rm split}$ is the number that we get after
    splitting all very massive stars into sub-clusters, as explained in section
    \ref{sec.shrinking}.
The reason why the last model has more stars than {\tt
    beta3b} at $T = 300$ is because it corresponds to a previous moment in the
    evolution and, as we explained above, protostars are allowed to merge with
    each other. This last case is a special one, and we ran a dedicated
    simulation for it. See section \ref{sec.ClusterInfall}.
    Also, we note that while the gas was originally distributed in a rather narrow radial range
    ($2a-5a$), we end up with stars even at distances $>100\,a$.  This is due to $N-$body
scattering, as many star particles are formed in relatively crowded regions of
the disc.
    \label{tab.nbody}
    }
\end{table}

\begin{figure*}
\resizebox{\hsize}{!}
          {\includegraphics[scale=1,clip]{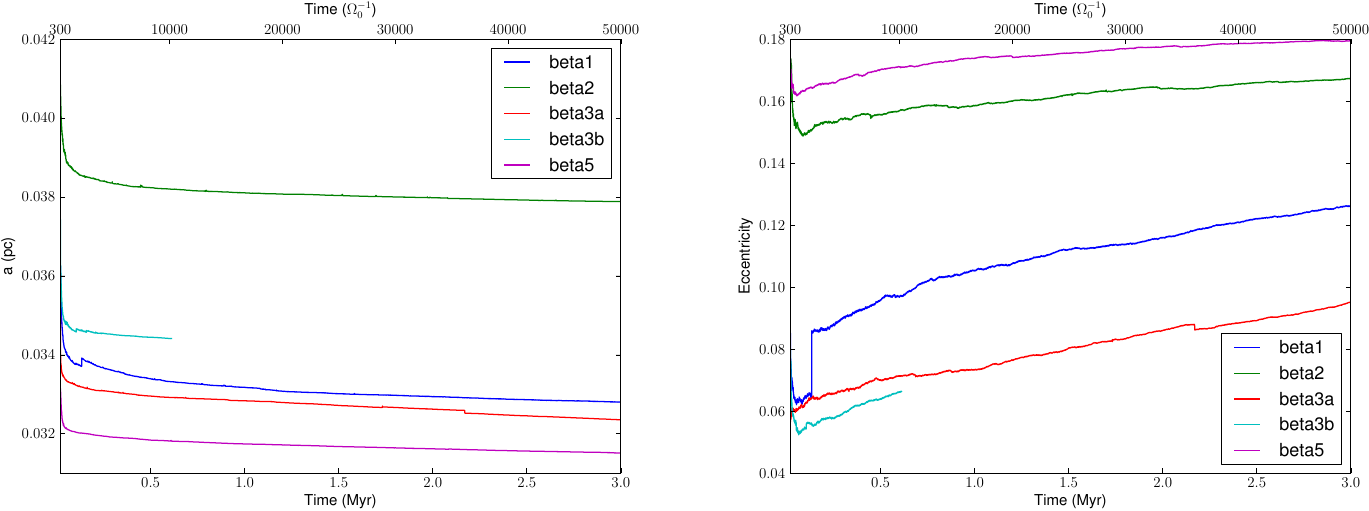}}
\caption
   {
{\em Left panel:} Evolution of the binary semi-major axis for all cases in the
$N-$body simulations in units of the initial orbital period $\Omega_{0}^{-1}$
and in Myrs. Case {\tt beta3b} was not integrated for more than 10,000 orbits because
of numerical issues due to the high
number of stars in that simulation. {\em Right panel:} Same for the eccentricity.
   }
\label{fig.EvolutionOrbElemNbody}
\end{figure*}

In figure \ref{fig.EvolutionOrbElemNbody} we see the evolution of $a$ and $e$
for all cases, integrated with {\sc Nbody6} with the results of the SPH
simulations as input parameter. After some 10,000 orbits the binaries reach a
stagnation point from which the decay becomes much slower.  The decay rates
$(1/a)(\Delta a/\Delta t)$ averaged over the time period from 0.5 Myr to 3 Myr
are: {\tt beta1} : $7.2\times 10^{-9} \,{\rm yr}^{-1}$, {\tt beta2} : $4.0\times
10^{-9} \,{\rm yr}^{-1}$, {\tt beta3a} : $8.0\times 10^{-9} \,{\rm yr}^{-1}$, and
{\tt beta5} : $4.0\times 10^{-9} \,{\rm yr}^{-1}$ (although for this case we
start at 1000 $\Omega_0^{-1}$, which means actually from 0.56 to 3.06 Myr).
In the first 0.1 Myr of the evolution, the significant drop in
semi-major axis corresponds to decay rates of $6.7 \times 10^{-7}\,{\rm
yr}^{-1}$ for {\tt beta1}, $5.4 \times 10^{-7}\,{\rm yr}^{-1}$ for {\tt beta2},
$5.4 \times 10^{-7}\,{\rm yr}^{-1}$ for {\tt beta3a} and $3.2 \times
10^{-7}\,{\rm yr}^{-1}$ for {\tt beta5}.

The early dynamical evolution (first few hundred $\Omega_0^{-1}$) is dominated
by close encounters between the MBH binary and stars on radial orbits (i.e. in
the loss cone of the binary). This is naturally accompanied by a high rate of
tidal disruptions (see figure \ref{fig.tidalmassall_time}) and a strong change
in orbital binding energy of the binary. In the following long-term evolution,
the loss cone has been depleted and the binary is subject to the secular
effects of the disk as a non-spherical background potential.  The effect of
this type of mass distribution is a slow exchange of orbital energy but a
rather efficient exchange of angular momentum \citep{MerrittVasiliev10}, which
is consistent with the significant increase in eccentricity that we observe in
this phase compared to the very slow decay rates in the semi-major axis.

\subsection{An infalling cluster of young stars}
\label{sec.ClusterInfall}

\begin{figure*}
\resizebox{\hsize}{!}
          {\includegraphics[scale=1,clip]{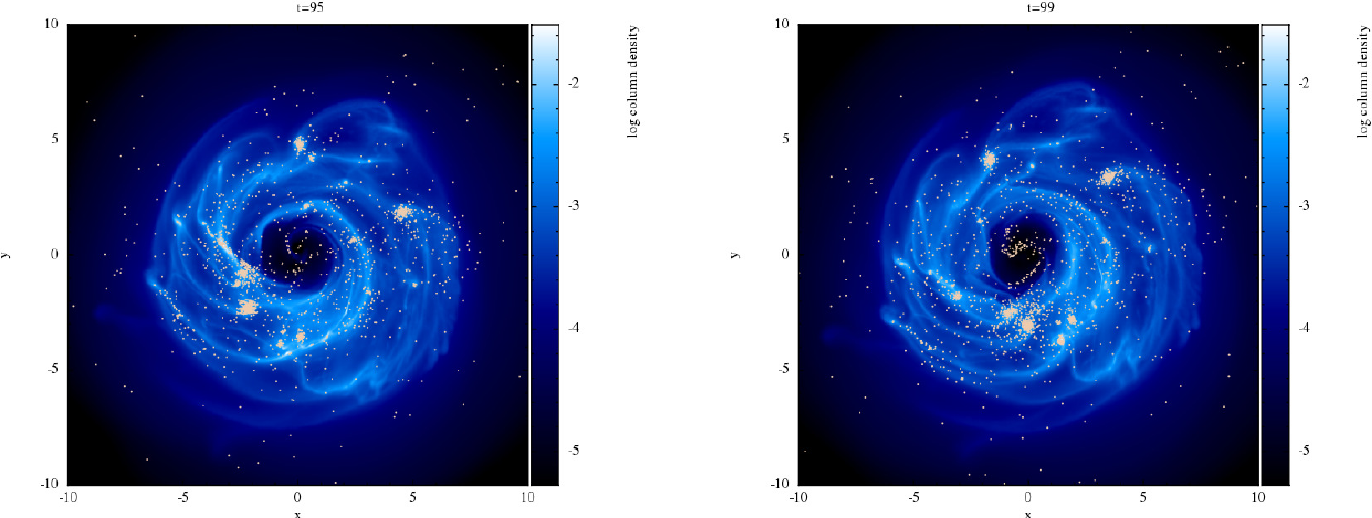}}
\caption
   {
Same as figure \ref{fig.GasDens_All} for the simulation {\tt beta3b}.
At the SPH time 95 the closest stellar cluster hurls itself on to the
binary and leads to an enhancement in the TDEs. We take the position of
stars and gas particles from the left panel to start a dedicated direct-summation
$N-$body integration which we name after this instant of time, {\tt beta3b95}.
   }
\label{fig.3b_mosaic}
\end{figure*}

 In the SPH simulations modelling disc fragmentation we see large
 amounts of stars falling to the immediate vicinity of the MBH binary.
 In particular, in simulation {\tt beta3b} we observe an infalling
 cluster
     \footnote{
As in the former footnote about the movie, from $T=95$ onwards in the simulation.
              }
at $T=95\,\Omega_{0}^{-1}$ (see figure \ref{fig.3b_mosaic}).

Since this is quite interesting from the stellar dynamics point of view, we run
a dedicated simulation for this particular situation with the direct-summation
code. Nonetheless, at this early stage in the evolution of the disc, there is a
significant mass in gas which has not yet transformed into stars.  If we ran
the simulation without taking into account the gas, the small stellar clusters
would dissolve, as their potential wells would be abruptly much shallower and the
stars could not be held together.  We therefore have to include a prescription
in the $N-$body simulations for the {role} of the gas, since including the gas
particles directly is well outside the scope of our work.

\begin{figure*}
\resizebox{0.92\hsize}{!}
          {\includegraphics[scale=1,clip]{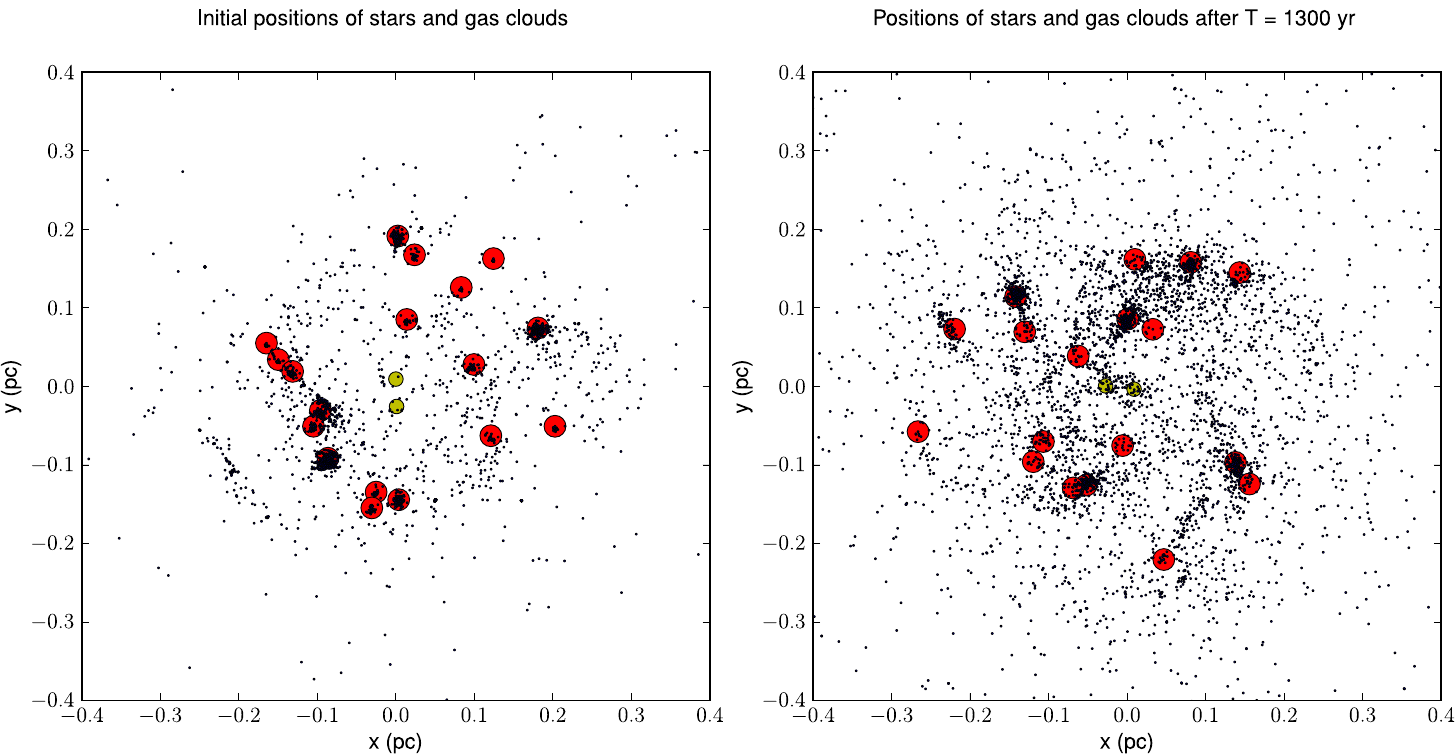}}
\caption
  {
{\em Left panel:} Initial configuration in the x-y plane perpendicular to the angular momentum vector of the
system for simulation {\tt
beta3b95} of table \ref{tab.nbody}.  Stars are shown with black dots, gas clouds
with red circles. The MBHs are depicted with green circles. {\em Right panel:}
The same system after the cluster falls on to the binary, after $\sim 1,300$
yrs. Note the enhanced number of stars in the vicinity of the binary. This
translates in a larger number of TDEs.
\label{fig.InfallingCluster}
   }
\end{figure*}

In this dedicated $N-$body simulation we model
each dense region that contains a non-negligible amount of mass as one particle
with a big softening length. For this, we define a sphere at every region of
interest. We then look at the SPH gas particle distribution and group together
all particles within this region, compute their total mass, center-of-mass
position and velocity and create one ``cloud particle'' with these properties
(see Fig. \ref{fig.InfallingCluster}).  In the subsequent $N-$body simulation
these particles are integrated separately, which required a modification in the
code.  In all gravitational interactions, the gravitational potential of the
cloud particle seen by a regular star is then computed as

\begin{equation}
\Phi_{\rm c} = - \frac{G M_{\rm c}}{r_{\rm c}+\epsilon},
\end{equation}

\noindent
where $M_{\rm c}$ and $r_{\rm c}$ are the mass and distance to the cloud particle and
$\epsilon$ denotes the softening length, taken to be of the order of the size
of the corresponding sub-cluster. Although the concept of cloud particles is
already implemented in the standard version of {\sc Nbody6}, we modified it to
integrate the cloud particles taking into account the gravitational potential of the
other clouds, stars and MBHs in order to follow correctly the orbits around the central
binary of MBHs.

The effect is that the particles in the sub-clusters now feel an additional
gravitational force corresponding to the cloud and thus stay within their
respective group for a longer time, which allows us to study their infall and
to analyse TDEs.  However, after one close encounter of a gas cloud with one of
the MBHs, the cloud would suffer a stripping from the cluster and now float
around as an unphysically big agglomeration of mass. This means that we can get
only a meaningful result for the very first encounter of each sub-cluster with
the binary. In this respect, when estimating the TDEs for the infalling
cluster, we will be giving a {\em lower limit}, since we cannot simulate
realistically further interactions of the cluster with the MBH
binary. In the right panel of figure \ref{fig.InfallingCluster} we show the
distribution of stars in the X--Y plane after the first interaction.

\section{Tidal disruption events}

During the direct-summation $N-$body runs any star entering the tidal
radius $R_{\rm T}$ of one of the MBHs is considered to be tidally
disrupted and its mass is added to the mass of the hole.  For a
solar-type star, this radius is \citep[see e.g.][for a derivation and
  examples]{Amaro-SeoaneLRR2012}

\begin{equation}
  R_{\rm T} = R_{\star}\, \left(\frac{M_{\rm BH}}{m}\right)^{1/3}.
  \label{eq:tidal}
\end{equation}

\noindent
In the last expression $M_{\rm BH}$ is the mass of {\it one} of the MBHs, $R_{\star}$ the
radius of the star and $m$ its mass.  In order to estimate the radius
of a star given its mass, we adopt the simple relation $R_{\star} \propto
m^{0.6}$ \citep{DemircanKahraman91,GordaSvechnikov1998}
with the normalization that a solar mass star has solar
radius. Using this in Eq. \ref{eq:tidal}, we can compute the tidal radius in
solar radii:

\begin{equation}
  R_{{\rm T},\odot} = 1.29\,m_{\odot}^{0.6}\, \left(\frac{M_{\rm BH}}{m}\right)^{1/3},
\end{equation}

\noindent
where $m_{\odot}$ is the mass of the star in solar masses  and the pre-factor comes
from an empirical fit for high-mass stars.

\begin{figure}
\resizebox{\hsize}{!}
          {\includegraphics[scale=1,clip]{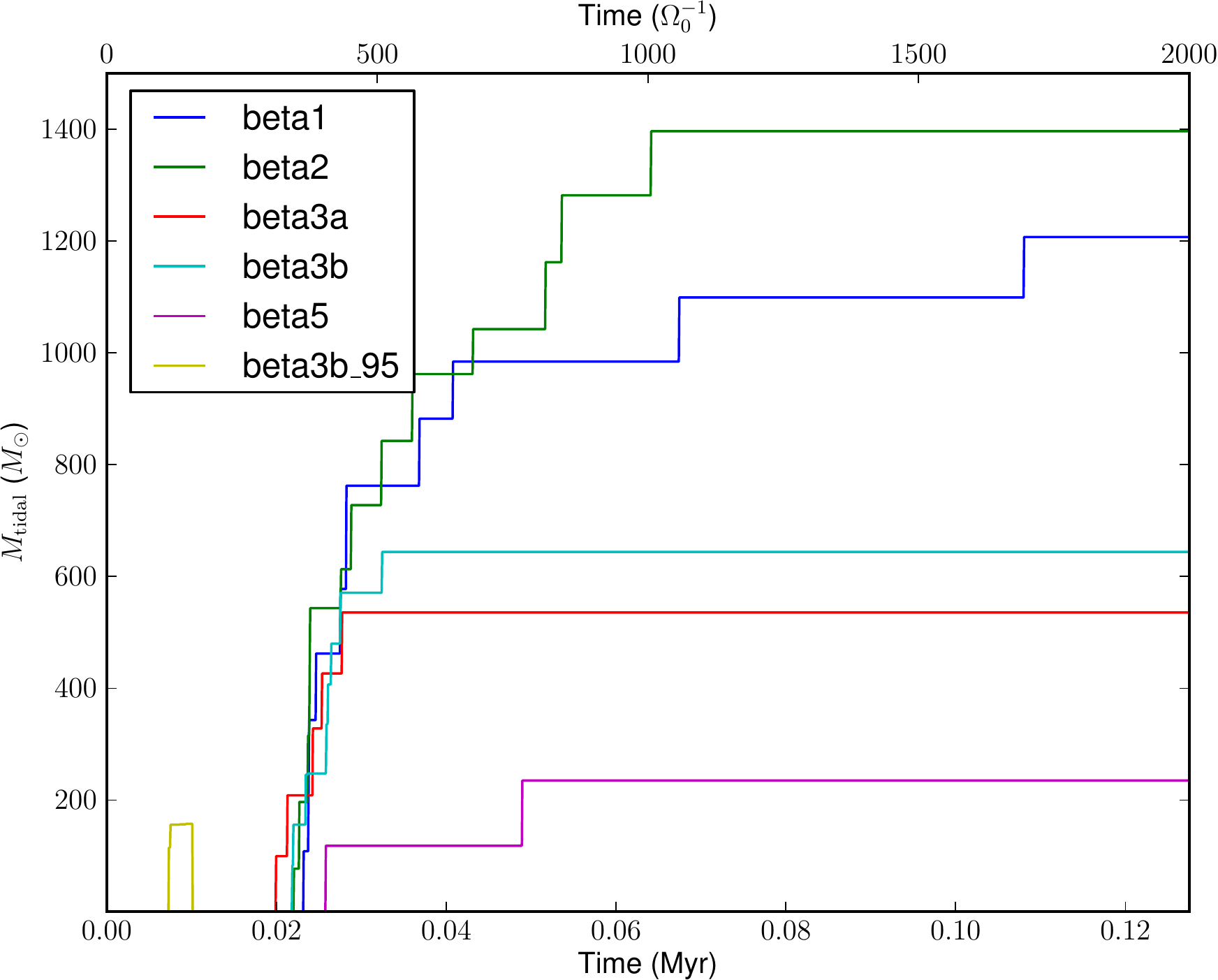}}
\caption
   {
Accumulated stellar mass in tidal disruptions for the different simulations of
table \ref{tab.nbody}.  Since we only track them in the $N-$body simulations,
the curves start at $T=300\,\Omega_0^{-1}$ but for model {\tt beta3b95}, which
as explained before, had a dedicated run. Because we cannot simulate
realistically more than one infall of the cluster, we stop it after the first
periapsis passage.
   }
\label{fig.tidalmassall_time}
\end{figure}

In figure \ref{fig.tidalmassall_time} we show the accumulated stellar mass
fraction in tidal disruptions for all simulations. Based on this figure and for
a time interval of 0.1 Myr after the initialization of the simulations, we can
convert the values in tidal disruption events, as shown in table
\ref{tab.TDEs}.  Notice that the rate is actually much higher at the beginning
of the simulations, but that result is likely to be affected by our
initialisation choices.

\begin{table}
\centering
    \begin{tabular}{|l|c|}
        Simulation   & TDEs (${\rm yr}^{-1}$) \\ \hline\hline
        {\tt beta1}  & $1.1\cdot10^{-4}$      \\
        {\tt beta2}  & $1.4\cdot10^{-4}$      \\
        {\tt beta3a}  & $6\cdot10^{-5}$        \\
        {\tt beta3b}  & $9\cdot10^{-5}$        \\
        {\tt beta5}  & $2\cdot 10^{-5}$       \\
        \hline
    \end{tabular}
\caption{
Tidal event rates for the simulations of table \ref{tab.nbody}.
\label{tab.TDEs}
}
\end{table}

\section*{Discussion}

In this work we have presented the first realisations of fragmenting discs
around a binary of two MBHs in SPH with star formation followed by
direct-summation $N-$body simulations of the resulting systems. We have
evaluated different fragmentation scenarios based on an approximation for the
cooling rate of the gaseous discs and different prescriptions for the
growth of protostars.

When the gas is almost completely depleted, we take the masses, positions and
velocities of the newly formed stars and feed them to the direct-summation
$N-$body integrations with the proviso that if the protostar has a mass above
$120\,M_{\odot}$, we convert it into an agglomeration of stars following a
Plummer profile of radius the Roche radius of the protostar to avoid
artificially-created very massive stars.

We find that the rate of decay in our direct $N-$body simulations is slower
than the $\approx 10^{-6}~{\rm yr}^{-1}$ found in the SPH simulations of
\cite{CuadraEtAl09}, when scaling to the same masses and separations.

We simulate with a dedicated direct-summation integration the particular case of
a simulation in which a cluster of stars that forms during the SPH simulation
falls on to the binary, the case {\tt beta3b95}. For this, we modify {\sc
Nbody6} to include ``gas cloud'' particles that allow the infalling cluster to
hold together in the dynamical simulation in which we cannot realistically
simulate the gas.

Infalling clusters such as this and the scattering of isolated stars lead to a
significant number of TDEs.  To make an accurate estimation, we made a second
modification of {\sc Nbody6} to implement stellar tidal disruptions, and we
find that the event rates lie between $2\cdot 10^{-5} - \sim 10^{-4}$ per
system per year, which lies on the high side of current (uncertain) estimates
for the TDE rate in standard galaxies, which typically lie between
$10^{-5}-10^{-6}\,{\rm yr}^{-1}$ \citep{Phinney1989,MT99,SU99}, and lie well within the observed
rates \citep{DonleyEtAl02,vanVelzenFarrar2012}. A
particular interesting signature of these TDEs is the ``reverberation mapping''
response of the circumbinary disc to a burst of emission produced by the TDEs.
The light from the burst excites the gas in the disc, producing emission lines.
The time-variability of the spectra, the {\em echo} of the TDE, during the
months after the burst could in principle allow us to constrain the disc
structure (Brem, Amaro-Seoane, Cuadra \& Komossa; part II of this paper to be
submitted).

\begin{figure}
\resizebox{\hsize}{!}
          {\includegraphics[scale=1,clip]{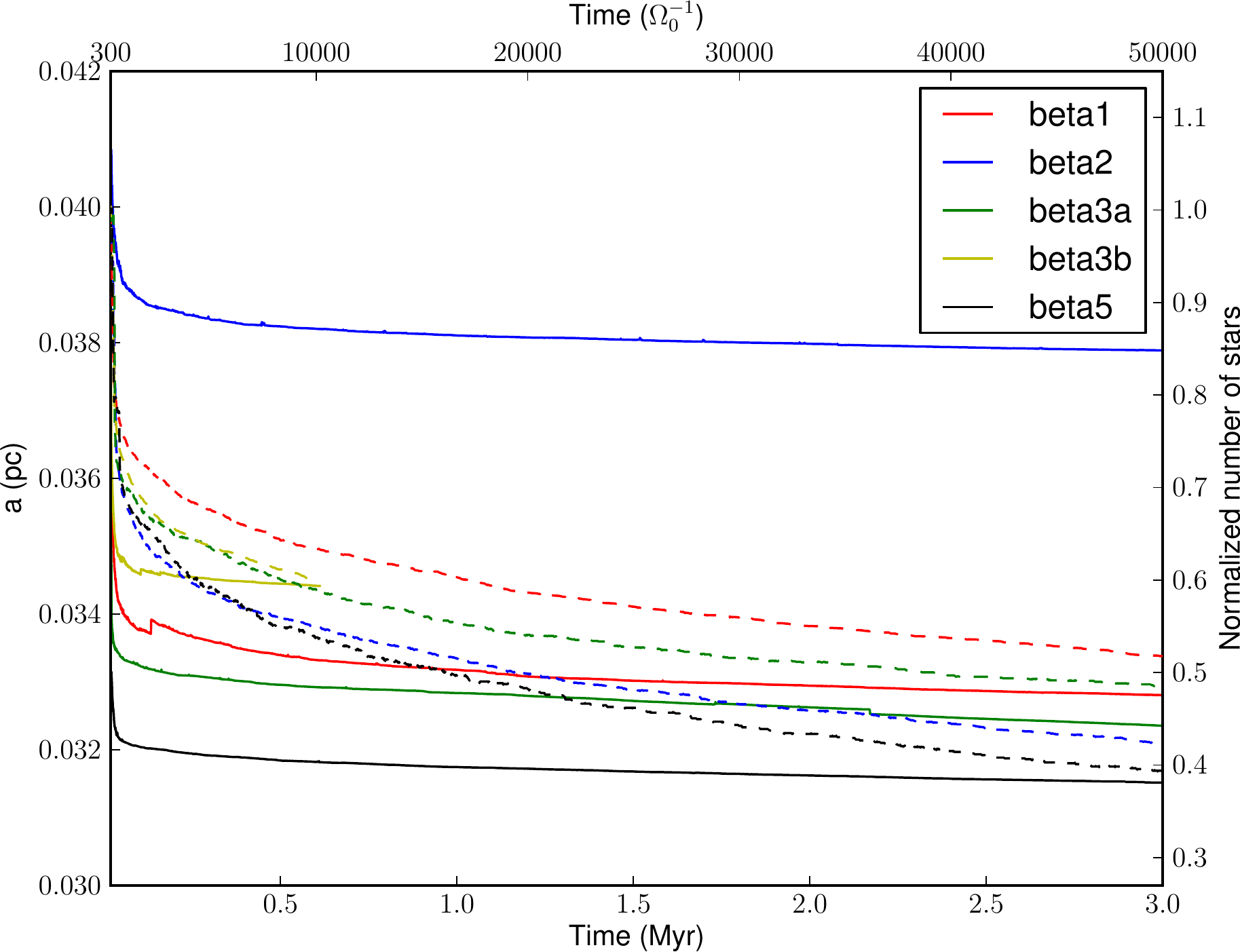}}
\caption
   {
Same as the left panel of figure \ref{fig.EvolutionOrbElemNbody} but including
the number of stars in each simulation, for the same colour but in dashed lines.
By the end of our numerical treatment we have lost at least 50\% in all cases.
   }
\label{fig.Semimajor_NumberStars}
\end{figure}

While our simulations cannot follow the evolution of the binary for much longer times, it is interesting to ask the
question whether the semi-major axis of the binary reaches distances that would
lead it to coalesce within a Hubble time because of the emission of
gravitational radiation, measurable in a LISA-like detector such as eLISA
\citep{Amaro-SeoaneEtAl2012}. For this, the binary has to shrink from an
initial semi-major of $a \approx 0.04\, {\rm pc}$ down to $a \approx 0.003
\,{\rm pc}$. This corresponds to an increase of orbital binding energy of about
one order of magnitude. The net change in binding energy after an interaction
with one bound star of mass $m_\star$ can be estimated as $\Delta E_\star =
{G m_\star M_{\rm bbh}}/{a}$.  We start the direct-summation simulations
with a ratio of stellar mass to MBH binary mass of $\approx 10\%$, so that ab
definitio the stellar mass that is formed is not enough for the binary to
shrink down to the phase in which the evolution is dominated by gravitational radiation. Indeed, if we consider all stars in the disc to be ejected, we estimate
in the limit of this low mass ratio that the total effect of the stellar disc
is of about $\delta E_{\rm tot} = {G q M^2_{\rm bbh}}/{a}$, where $q$ is
the mass ratio of stellar mass to BH mass. Following an argument similar to
e.g.  \cite{Quinlan96,SesanaEtAl07}, if we compare this to the orbital energy
at semi-major axis $a$, one finds that the relative change after ejecting all
the stars is $\delta E_{\rm tot}/E \approx q/\nu$, where $\nu = 0.2$ is the
symmetric mass ratio, well below what would be necessary to shrink the binary
by one order of magnitude. We note that indeed ejecting half of the stellar
mass only shrinks the binary semi-major axis by $< 25\%$, as we see in the
first 3 Myr of our $N-$body simulations, in figure
\ref{fig.Semimajor_NumberStars}.  While this is true for our specific scenario,
we note that further episodes of gas inflow towards the centre could
potentially trigger more episodes of star formation in the disc, which would
lead to star scattering and a further skrinkage.
Moreover, while we have focused
on the effect of stars formed in-situ on the binary, but the system will
be surrounded by a stellar cusp that constitutes an additional source of
shrinkage for the binary. The supply of stars that will interact with it can be enhanced by additional
mechanisms in a more realistic picture than that of an isolated, spherically
symmetric galactic nucleus, as we discussed in the introduction.

\acknowledgments

PAS is indebted to the Universidad Cat{\'o}lica for their hospitality and
support for a two-week visit, as well as to Jura Borissova and her group from
Universidad Valpara{\'\i}so for a short visit, and to Cristi{\'a}n Maureira,
for his help and discussions on the way to Valpara{\'\i}so.  JC acknowledges
support from FONDAP (15010003), FONDECYT (11100240), Basal (PFB0609), VRI-PUC
(Inicio 16/2010) and the European Commission's Framework Programme 7 through
the Marie-Curie IRSES project LACEGAL (PIRSES-GA-2010-269264). This work was
supported in  part by the National Science Foundation under Grant No.\ 1066293
and the hospitality of the Aspen Center for Physics and by the Transregio 7
``Gravitational Wave Astronomy'' financed by the Deutsche
Forschungsgemeinschaft DFG (German Research Foundation).

\label{lastpage}
\end{document}